\documentclass[12pt]{article}

\usepackage{amsmath}
\usepackage{amssymb}
\usepackage{amscd}
\usepackage{mathtools}
\usepackage{mathrsfs}
\usepackage{physics}
\usepackage{bm}
\usepackage[pdftex]{graphicx}
\pdfoutput=1
\usepackage{ytableau}
\usepackage[margin =27truemm]{geometry}
\usepackage[hang,small,bf]{caption}
\usepackage[subrefformat=parens]{subcaption}
\usepackage{cite}
\begin{document}
\setlength{\baselineskip}{0.6cm}

\begin{titlepage}
\begin{flushright}
NITEP 207
\end{flushright}

\vspace*{10mm}%

\begin{center}{\Large\bf
New Models of $SU(6)$ Grand Gauge-Higgs Unification
}
\end{center}
\vspace*{10mm}
\begin{center}
{\large Nobuhito Maru}$^{a,b}$ and
{\large Ryujiro Nago}$^{a}$ 
\end{center}
\vspace*{0.2cm}
\begin{center}
${}^{a}${\it
Department of Physics, Osaka Metropolitan University, \\
Osaka 558-8585, Japan}
\\
${}^{b}${\it Nambu Yoichiro Institute of Theoretical and Experimental Physics (NITEP), \\
Osaka Metropolitan University,
Osaka 558-8585, Japan}
\end{center}

\vspace*{20mm}


\begin{abstract}
A five dimensional $SU(6)$ grand gauge-Higgs unification compactified on $S^1/Z_2$ is discussed. 
We propose new sets of the $SU(6)$ representations where the quarks and leptons in one generation 
are embedded and there is no extra massless exotic fermions absent in the Standard Model. 
The correct electroweak symmetry breaking pattern can be realized by introducing some adjoint fermions. 
We also analyze whether a viable Higgs mass can be obtained. 
\end{abstract}

\end{titlepage}

\section{Introduction}
Gauge-Higgs unification (GHU) is one of the scenarios solving the hierarchy problem in the Standard Model (SM) 
without relying on supersymmetry \cite{Manton, Fairlie, Hosotani1, Hosotani2}.  
In this scenario, the SM Higgs field is identified with the extra component of the higher dimensional gauge field. 
From this construction, the interactions in the gauge and Higgs sectors are controlled by the gauge symmetry. 
As a result, the Higgs potential and Higgs mass is generated at quantum level and becomes finite 
thanks to the gauge symmetry \cite{HIL, SSS, ABQ, MY, HMTY, LMH, HLM}. 
Reminding that the hierarchy problem was originally discussed in the grand unified theory (GUT), 
it is natural to extend GHU to GUT and important to study the viability of the GUT extended GHU, 
 which we call the grand gauge-Higgs unification (GGHU). 
Along this line of thought, there has been many studies on GGHU in various directions 
\cite{LM, KTY, HosotaniGGHU, MYT, ABBGW}. 

We discuss a five dimensional (5D) $SU(6)$ GGHU compactified on an orbifold $S^1/Z_2$. 
This model has been discussed in \cite{LM} where the sets of representations 
including only quarks and leptons per generation as massless fields was proposed, 
namely those without exotic massless fermions absent in the SM.  
The proposal is remarkable in that the 4D gauge anomaly is trivially cancelled 
and there are no exotic massless fermions which usually appear 
after the decomposition of the higher rank representations in GUT. 
However, the right-handed down quark and charged lepton are embedded into a different representations 
having the left-handed quark and lepton doublets. 
This implies that the down-type quark and the charged lepton Yukawa couplings cannot be generated 
 through the gauge interaction in the 5D bulk, which leads us to the construction of Yukawa coupling 
 by an interplay between bulk and brane localized fermions \cite{SSS, MYT, ABBGW}.  

In this paper, we propose new sets of $SU(6)$ representations 
including only quarks and leptons per generation and no exotic fermions. 
Performing systematic group theoretical considerations, 
we find three type sets of $SU(6)$ representations satisfying the above mentioned properties. 
One of them has been already discussed in \cite{LM}, namely $\bm{6}^* \oplus \bm{6'}^* \oplus \bm{20}$.  
Other two types, $\bm{15} \oplus \bm{15'} \oplus \bm{1}$ and 
$\bm{15} \oplus \bm{21} \oplus \bm{1}$, are our new results. 
Using combinations of these sets of fermions, we can obtain three generation of quarks and leptons. 
Furthermore, one-loop Higgs potential is calculated with various combinations of the representation sets  
for three generations and the extra fermions in the adjoint representation. 
It will be shown that the electroweak symmetry breaking can be realized with some adjoint fermions. 
and the experimental value of Higgs mass can be obtained by adding adjoint fermions further.  

This paper is organized as follows. 
In the next section, we introduce the gauge and Higgs sector of 5D $SU(6)$ GGHU. 
In section 3, we study the possible sets of $SU(6)$ representations 
including only quarks and leptons per generation as massless fields. 
Higgs potential analysis in our model is performed in section 4. 
The possibility of electroweak symmetry breaking and a viable Higgs mass are discussed. 
Section 5 is devoted to conclsions of our paper. 
In Appendix A, the classification of the boundary conditions is discussed. 
A systematic group theoretical search for the $SU(6)$ representation sets 
without exotic fermions is provided in Appendix B. 

\section{5D $SU(6)$ grand gauge-Higgs unification compactified on $S^1/Z_2$
}
In this section, we consider a 5D $SU(6)$ GGHU 
 where an extra space $y$ is compactified on $S^1/Z_2$ with a radius $R$. 
Due to such a compactification, 
 the boundary conditions of the 5D fields have to be given. 
As for $S^1$, the (anti-)periodic boundary condition is imposed. 
As for $Z_2$, the $Z_2$ parity transformations on each fixed point $y=0, \pi R$ have to be imposed.   
%
%
4D gauge fields $A_{\mu} \, ( \mu = 0, \, 1, \, 2, \, 3)$ and the fifth component $A_5$ are transformed as
\begin{equation}
	\begin{alignedat}{3}
		A_{\mu} &\to \hat{P} A_{\mu} \hat{P}^{\dag}, \quad &A_5 &\to -\hat{P} A_5 \hat{P}^{\dag} && \qq{at} y = 0 \, , \\
		A_{\mu} &\to \hat{P'} A_{\mu} \hat{P'}^{\dag},		&A_5 &\to -\hat{P'} A_5 \hat{P'}^{\dag} && \qq{at} y = \pi R \, ,
	\end{alignedat}
\end{equation}
where $\hat{P}, \hat{P'}$ are $6 \times 6$ parity matrices given by (\ref{paritymatrices}).  
A 5D field $\Phi(x,y)$ has eigenvalues of $\hat{P}$ and $\hat{P'}$ specified as $( P, P')$, 
 and we can expand $\Phi(x,y)$ in terms of Kaluza-Klein (KK) modes correspondingly. 
\begin{equation}
	\begin{aligned}
		\Phi^{(+,+)}(x,y) &= \frac{1}{\sqrt{2\pi R}}\phi_0(x) + \frac{1}{\sqrt{\pi R}}\sum_{n=1}^{\infty} \phi_n^{(+,+)}(x)\cos \qty(\frac{n}{R}\,y), \\
		\Phi^{(+,-)}(x,y) &= \frac{1}{\sqrt{\pi R}}\sum_{n = 0}^{\infty} \phi_n^{(+,-)}(x)\cos \qty(\frac{n + 1/2}{R}\,y), \\
		\Phi^{(-,+)}(x,y) &= \frac{1}{\sqrt{\pi R}}\sum_{n = 0}^{\infty} \phi_n^{(-,+)}(x)\sin \qty(\frac{n + 1/2}{R}\,y), \\
		\Phi^{(-,-)}(x,y) &= \frac{1}{\sqrt{\pi R}}\sum_{n = 0}^{\infty} \phi_n^{(-,-)}(x)\sin \qty(\frac{n}{R}\,y).
	\end{aligned}
\end{equation}
We note that only $\Phi^{(+,+)}(x,y)$ has a 4D massless zero mode $\phi_0(x)$.
Next, we consider 
a gauge symmetry breaking of $SU(6)$ to $SU(3)_c \times SU(2)_L \times U(1)_Y \times U(1)_X$ 
by $Z_2$ boundary conditions ($\hat{P}$, $\hat{P'}$).
As we will show in Appendix A, it is sufficient to consider the following $\hat{P}$ and $\hat{P'}$
to obtain the above symmetry breaking. 
%
 \begin{equation}
	\begin{aligned}
		\hat{P} = \; & \mathrm{diag}( +1, +1, +1, \, +1, +1, \, -1) \qq{at} y = 0, \\
		\hat{P'} = \; & \mathrm{diag} (+1, +1, +1, \, -1, -1, \, +1) \qq{at} y = \pi R .
	\end{aligned}
	\label{paritymatrices}
\end{equation}
%
The generator of $U(1)_X$, $T_{U(1)_X} = \frac{1}{\sqrt{62}} \mathrm{diag}(1,1,1,1,1,-5)$ 
corresponds to one of generators of $SU(6)$.
As for the extra $U(1)_X$ gauge symmetry, it can be spontaneously broken 
if we introduce a $U(1)_X$ charged 4D scalar field and the double-well potential localized on the $y=0$ brane, 
where the $SU(5) \times U(1)$ gauge symmetry is unbroken. 

The $Z_2$ parity assignments of matrix elements of $A_{\mu , 5}$ are explicitly written as
\begin{equation}
	\begin{aligned}
		A_{\mu} &= 
			\left(
			\begin{array}{ccc|cc|c}
				(+,+) & (+,+) & (+,+) & (+,-) & (+,-) & (-,+) \\ 
				(+,+) & (+,+) & (+,+) & (+,-) & (+,-) & (-,+) \\
				(+,+) & (+,+) & (+,+) & (+,-) & (+,-) & (-,+) \\[2pt] \hline \vspace{-10pt}
				{} & {} & {} & {} & {} & {} \\ 
				(+,-) & (+,-) & (+,-) & (+,+) & (+,+) & (-,-) \\
				(+,-) & (+,-) & (+,-) & (+,+) & (+,+) & (-,-) \\[2pt] \hline \vspace{-10pt}
				{} & {} & {} & {} & {} & {} \\
				(-,+) & (-,+) & (-,+) & (-,-) & (-,-) & (+,+) 
			\end{array}
			\right) ,\\
		A_{5} &= 
			\left(
			\begin{array}{ccc|cc|c}
				(-,-) & (-,-) & (-,-) & (-,+) & (-,+) & (+,-) \\
				(-,-) & (-,-) & (-,-) & (-,+) & (-,+) & (+,-) \\
				(-,-) & (-,-) & (-,-) & (-,+) & (-,+) & (+,-) \\[2pt] \hline \vspace{-10pt}
				{} & {} & {} & {} & {} & {} \\
				(-,+) & (-,+) & (-,+) & (-,-) & (-,-) & (+,+) \\
				(-,+) & (-,+) & (-,+) & (-,-) & (-,-) & (+,+) \\[2pt] \hline \vspace{-10pt}
				{} & {} & {} & {} & {} & {} \\
				(+,-) & (+,-) & (+,-) & (+,+) & (+,+) & (-,-) 
			\end{array}
			\right).
	\end{aligned}
\end{equation}
We see in $A_5$ matrix elements that the doublet-triplet mass splitting in Higgs sector is realized 
 by $Z_2$ boundary conditions, namely the SM Higgs doublet is massless but the colored Higgs triplet has no massless mode.


\section{Types of $SU(6)$ model}
In this section, we focus on a fermion sector in our model. 
As in the gauge fields, we provide the $Z_2$ parity transformations for fermions as 
\begin{equation}
\label{fermiontr.}
	\begin{alignedat}{2}
		\psi_{\mathcal{R}} &\to - \eta_{\mathcal{R}} \hat{P}(\mathcal{R}) \gamma^5 \psi_{\mathcal{R}} && \qq{at} y = 0 \, , \\
		\psi_{\mathcal{R}} &\to - \eta'_{\mathcal{R}} \hat{P'}(\mathcal{R}) \gamma^5 \psi_{\mathcal{R}} && \qq{at} y = \pi R. 
	\end{alignedat}
\end{equation}
$\hat{P}(\mathcal{R})$ and $\hat{P'}(\mathcal{R})$ are defined as $Z_2$ parity matrices 
 acting on the fermion $\psi_{\mathcal{R}}$ in an irreducible representation $\mathcal{R}$ 
 at the fixed points $ y = 0 $ and $y = \pi R $, respectively. 
 $\gamma^5$ is an ordinary gamma matrix defined by 4D gamma matrices 
  as $\gamma^5 = i \gamma^0\gamma^1\gamma^2\gamma^3$.  
$\eta_{\mathcal{R}}$ and $\eta'_{\mathcal{R}} $ take a value of $+1$ or $-1$ in each representation. 
For example, $\hat{P} \coloneqq \hat{P}(\bm{6})$ and $\hat{P'} \coloneqq \hat{P'}(\bm{6})$ are $6 \times 6$ matrices.

As a guiding principle, let us impose one condition to constrain a fermion matter content of our model.
\\[5pt]
\quad ($\star$) \quad All representations introduced to obtain quarks and leptons contain no exotic fermions.
\\[5pt]
Here, the exotic fermions mean massless fermions, which have the SM group representations 
 but are not included in the SM.  

Which representations of $SU(6)$ satisfy the condition ($\star$) ?
In other words, we may ask which representations contain exotic fermions. 
To answer this question, we consider a fifty six dimensional representation $\bm{56}$ as an example.
$\bm{56}$ is a three-rank totally symmetric tensor of $SU(6)$ 
 and corresponds to the following Young tableau
\begin{equation*}
	\begin{ytableau}
		{} & {} & {} 
	\end{ytableau}
	\, .
\end{equation*}
Decomposing $\bm{56}$ into representations of $SU(3)_c \times SU(2)_L \times U(1)_Y \times U(1)_X$ and
focusing on blocks as
\begin{equation*}
	\begin{aligned}
		\begin{ytableau}
			{} & {} & \none & a 
		\end{ytableau}
		\, ,
	\end{aligned}
\end{equation*}
one finds that $\bm{56}$ has the fields $ \psi_{\alpha}, \, \psi_{\beta} $ with following representations
\begin{equation}
\label{56}
	\begin{aligned}
	\ytableausetup{ boxsize = 1.3em }
		&\psi_{\alpha} : \qty( \; 
		\raisebox{-1mm}{$
			\begin{ytableau}
				{} & {} & a 
			\end{ytableau}$}\: , \:
			\bullet \: , \:
			\bullet\:) \, , \\
		&\psi_{\beta} : \qty( \; 
		\raisebox{-1mm}{$
			\begin{ytableau}
				{} & {} 
			\end{ytableau}$} \: , \:
			\raisebox{-1mm}{$
			\begin{ytableau}
				a
			\end{ytableau}$} \: , \:
			\bullet \: ) \, ,
	\end{aligned}
\end{equation}
where the first, the second and the third boxes in parenthesis are representations 
 under the $SU(3)_c$, $SU(2)_L$ and $U(1)_X$ respectively and the $\bullet$ means the trivial representation. 
We note that these representations are not present in SM.
$Z_2$ parities $(P, P')$ of ${\psi_{\alpha}}_L \, , \, {\psi_{\alpha}}_R \, ,\, {\psi_{\beta}}_L $ and ${\psi_{\beta}}_R$ 
can be read from (\ref{paritymatrices})
%
\begin{equation}
	\begin{alignedat}{3}
		{\psi_{\alpha}}_L :\, &\big( &   & \eta_{\bm{56}}, &	  & \eta_{\bm{56}}' \, \big) \, , \\
		{\psi_{\alpha}}_R :\, &\big( &  -& \eta_{\bm{56}}, & \; -& \eta_{\bm{56}}' \, \big) \, ,\\
		{\psi_{\beta}}_L :\, &\big( &   & \eta_{\bm{56}}, & 	 -& \eta_{\bm{56}}' \, \big) \, , \\
		{\psi_{\beta}}_R :\, &\big( & -& \eta_{\bm{56}}, &  	  & \eta_{\bm{56}}' \, \big),
	\end{alignedat}
\end{equation}
where $L$ and $R$ denote 4D chiralities defined by $\gamma^5=-1$ and $1$, respectively. 
Choosing any kind of $Z_2$ parity $(\eta_{\bm{56}}, \eta'_{\bm{56}})$, one of them necessarily has a massless zero mode.
That is, $\bm{56}$ contains an exotic fermion absent in the SM and 
does not satisfy the condition ($\star$).
We apply the same method to all irreducible representations in $SU(6)$ 
 and find that representations satisfying the condition ($\star$) are only
\begin{equation}
	\bm{6}, \bm{6}^*, \bm{15}, \bm{15}^*, \bm{20}, \bm{21}, \bm{21}^*.
\end{equation}
The detail derivation of this result is shown in Appendix B. 

Now, we determine a matter content of fermions of our model. 
Since the representation $\mathcal{R}^*$ is equivalent to the complex conjugate of $\mathcal{R}$
as long as we consider all choices of $(\eta_{\mathcal{R}^*}, \, \eta'_{\mathcal{R}^*} )$,
we discuss only $\bm{6}^*, \bm{15}, \bm{20}, \bm{21}$ from now on.
The decomposition of these representations into $SU(3) \times SU(2) \times U(1)_Y \times U(1)_X$ is given by
\begin{equation}
\label{representations}
	\begin{aligned}
		\bm{6}^* &= 
			\left\{
				\begin{aligned}
					&\bm{6}^*_L 
						= (\bm{3}^*, \bm{1})^{ ( -\eta_{\bm{6}^*}, \, \eta'_{\bm{6}^*} ) }_{ (1/3, -1) }
							\oplus (\bm{1}, \bm{2})^{ ( -\eta_{\bm{6}^*}, \, -\eta'_{\bm{6}^*} ) }_{ ( -1/2, -1) }
							\oplus (\bm{1}, \bm{1})^{ ( \eta_{\bm{6}^*}, \, \eta'_{\bm{6}^*} ) }_{ (0, 5) } \\[2mm]
					&\bm{6}^*_R 
						= (\bm{3}^*, \bm{1})^{ ( \eta_{\bm{6}^*}, \, -\eta'_{\bm{6}^*} ) }_{ (1/3, -1) }
							\oplus (\bm{1}, \bm{2})^{ ( \eta_{\bm{6}^*}, \, \eta'_{\bm{6}^*} ) }_{ ( -1/2, -1) }
							\oplus (\bm{1}, \bm{1})^{ ( -\eta_{\bm{6}^*}, \, -\eta'_{\bm{6}^*} ) }_{ (0, 5) } 
				\end{aligned}
			\right. \\[2mm]
		\bm{15} &= 
			\left\{
				\begin{aligned}
					\bm{15}_L &= 
						(\bm{3}, \bm{1})^{ ( -\eta_{\bm{15}}, \, \eta'_{\bm{15}} ) }_{ (-1/3, -4) } 
							\oplus(\bm{1}, \bm{2})^{ ( -\eta_{\bm{15}}, \, -\eta'_{\bm{15}} ) }_{ (1/2, -4) } 
							\oplus(\bm{3}^*, \bm{1})^{ ( \eta_{\bm{15}}, \, \eta'_{\bm{15}} ) }_{ (-2/3, 2 ) } \\ & \hspace{3cm}
							\oplus(\bm{3}, \bm{2})^{ ( \eta_{\bm{15}}, \, -\eta'_{\bm{15}} ) }_{ ( 1/6, 2 ) }
							\oplus(\bm{1} ,\bm{1})^{ ( \eta_{\bm{15}}, \, \eta'_{\bm{15}} ) }_{ ( 1, 2 ) } \\[2mm]
					\bm{15}_R &=
						(\bm{3}, \bm{1})^{ ( \eta_{\bm{15}}, \, -\eta'_{\bm{15}} ) }_{ (-1/3, -4) } 
							\oplus(\bm{1}, \bm{2})^{ ( \eta_{\bm{15}}, \, \eta'_{\bm{15}} ) }_{ (1/2, -4) } 
							\oplus(\bm{3}^*, \bm{1})^{ ( -\eta_{\bm{15}}, \, -\eta'_{\bm{15}} ) }_{ (-2/3, 2 ) } \\ &\hspace{3cm}
							\oplus(\bm{3}, \bm{2})^{ ( -\eta_{\bm{15}}, \, \eta'_{\bm{15}} ) }_{ ( 1/6, 2 ) }
							\oplus(\bm{1} ,\bm{1})^{ ( -\eta_{\bm{15}}, \, -\eta'_{\bm{15}} ) }_{ ( 1, 2 ) } 
				\end{aligned}
			\right. \\[2mm]
		\bm{20} &= 
			\left\{ 
				\begin{aligned}
					\bm{20}_{L} & = \,
						(\bm{3}, \bm{2})^{ ( -\eta_{\bm{20}}, \, -\eta'_{\bm{20}} ) }_{ ( 1/6, -3 ) }
							\oplus(\bm{3}^*, \bm{1})^{ ( -\eta_{\bm{20}}, \, \eta'_{\bm{20}} ) }_{ ( -2/3, -3 ) }	
							\oplus(\bm{1}, \bm{1})^{ ( -\eta_{\bm{20}}, \, \eta'_{\bm{20}} ) }_{ ( 1, -3 ) } \\
							& \hspace{3cm}
							\oplus(\bm{3}^*, \bm{2})^{ ( \eta_{\bm{20}}, \, -\eta'_{\bm{20}} ) }_{ ( -1/6, 3 ) }	
							\oplus(\bm{3}, \bm{1})^{ ( \eta_{\bm{20}}, \, \eta'_{\bm{20}} ) }_{ ( 2/3, 3 ) }	
							\oplus(\bm{1}, \bm{1})^{ ( \eta_{\bm{20}}, \, \eta'_{\bm{20}} ) }_{ ( -1, 3 ) } \\[2mm]
					\bm{20}_{R}  &= \,
						(\bm{3}, \bm{2})^{ ( \eta_{\bm{20}}, \, \eta'_{\bm{20}} ) }_{ ( 1/6, -3 ) }
							\oplus(\bm{3}^*, \bm{1})^{ ( \eta_{\bm{20}}, \, -\eta'_{\bm{20}} ) }_{ ( -2/3, -3 ) }	
							\oplus(\bm{1}, \bm{1})^{ ( \eta_{\bm{20}}, \, -\eta'_{\bm{20}} ) }_{ ( 1, -3 ) } \\
							& \hspace{3cm}
							\oplus(\bm{3}^*, \bm{2})^{ ( -\eta_{\bm{20}}, \, \eta'_{\bm{20}} ) }_{ ( -1/6, 3 ) }	
							\oplus(\bm{3}, \bm{1})^{ ( -\eta_{\bm{20}}, \, -\eta'_{\bm{20}} ) }_{ ( 2/3, 3 ) }	
							\oplus(\bm{1}, \bm{1})^{ ( -\eta_{\bm{20}}, \, -\eta'_{\bm{20}} ) }_{ ( -1, 3 ) }
				\end{aligned}
			\right. \\[2mm]
		\bm{21} &= 
			\left\{
				\begin{aligned}
					\bm{21}_L &= 
						(\bm{6}, \bm{1})^{ ( \eta_{\bm{21}}, \, \eta'_{\bm{21}} ) }_{ (-2/3, 2) } 
							\oplus(\bm{3}, \bm{2})^{ ( \eta_{\bm{21}}, \, -\eta'_{\bm{21}} ) }_{ ( 1/6, 2) } 
							\oplus(\bm{3}, \bm{1})^{ ( -\eta_{\bm{21}}, \, \eta'_{\bm{21}} ) }_{ (-1/3, -4 ) } \\ & \hspace{3cm}
							\oplus(\bm{1}, \bm{3})^{ ( \eta_{\bm{21}}, \, \eta'_{\bm{21}} ) }_{ ( 1, 2 ) }
							\oplus(\bm{1} ,\bm{2})^{ ( -\eta_{\bm{21}}, \, -\eta'_{\bm{21}} ) }_{ ( 1/2, -4 ) }
							\oplus(\bm{1} ,\bm{1})^{ ( \eta_{\bm{21}}, \, \eta'_{\bm{21}} ) }_{ ( 0, -10 ) } \\[2mm]
					\bm{21}_R &= 
						(\bm{6}, \bm{1})^{ ( -\eta_{\bm{21}}, \, -\eta'_{\bm{21}} ) }_{ (-2/3, 2) } 
							\oplus(\bm{3}, \bm{2})^{ ( -\eta_{\bm{21}}, \, \eta'_{\bm{21}} ) }_{ ( 1/6, -4) } 
							\oplus(\bm{3}, \bm{1})^{ ( \eta_{\bm{21}}, \, -\eta'_{\bm{21}} ) }_{ (-1/3, -4 ) } \\ & \hspace{3cm}
							\oplus(\bm{1}, \bm{3})^{ ( -\eta_{\bm{21}}, \, -\eta'_{\bm{21}} ) }_{ ( 1, 2 ) }
							\oplus(\bm{1} ,\bm{2})^{ ( \eta_{\bm{21}}, \, \eta'_{\bm{21}} ) }_{ ( 1/2, -4 ) }
							\oplus(\bm{1} ,\bm{1})^{ ( -\eta_{\bm{21}}, \, -\eta'_{\bm{21}} ) }_{ ( 0, -10 ) } 
				\end{aligned}
			\right. \\[2mm]
	\end{aligned}
\end{equation}
where the bold face numbers in the right-hand side are the representations under $SU(3)_c \times SU(2)_L$. 
The numbers in the subscript are the charges of $U(1)_Y \times U(1)_{X} $.
Embedding of quarks and leptons into $SU(6)$ representations is summarized in Table 1.
\begin{table}[h]
\centering
	\begin{tabular}{c|c|c|c|c} \hline
		{} 	      & $\bm{6}^*$ & $\bm{15}$ & $\bm{20}$ & $\bm{21}$ \\ \hline \hline
		$ ( \eta_{\mathcal{R}}, \, \eta'_{\mathcal{R}} ) = (+,+)$ & $\times$ & $u_R , \,l_L, \, e_R$ & $\times$ &$\times$ \\ \hline
		$ ( \eta_{\mathcal{R}}, \, \eta'_{\mathcal{R}} ) = (+,-)$ & $\times$ & $q_L, \, d_R$ & $\times$ & $q_L, \, d_R$  \\ \hline
		$ ( \eta_{\mathcal{R}}, \, \eta'_{\mathcal{R}} ) = (-,+)$ & $d_R$ & $\times$ & $\times$ & $\times$ \\ \hline
		$ ( \eta_{\mathcal{R}}, \, \eta'_{\mathcal{R}} ) = (-,-)$ & $l_L , \, \nu_R$ & $\times$ & $q_L, \, u_R, \, e_R$ & $\times$ \\ \hline
	\end{tabular}
	\caption{Embedding of quarks and leptons into $SU(6)$ representations.}
\end{table} 
\\
In this table, it is shown that which kind of massless fermions are included in each representation and $Z_2$ parity.
A symbol $\times$ means that the corresponding representation has exotic fermions.
Looking at this table, we can construct a set of representations 
containing one generation of quarks and leptons with no exotic fermions.
Let us give an examaple.
Taking $ ( \eta_{\bm{6}^*}, \, \eta'_{\bm{6}^*} ) = (-,+)$, $d_R$ is obtained from $\bm{6}^*$.
Next, considering another fundamental representation $\bm{6}'^*$ with $ ( \eta_{\bm{6'}^*}, \, \eta'_{\bm{6'}^*} ) = (-,-)$, 
$l_L$ and $\nu_R$ are obtained.
Furthermore, taking $ ( \eta_{\bm{20}}, \, \eta'_{\bm{20}} ) = (-,-)$, 
$q_L$, $u_R$ and $e_R$ are obtained from $\bm{20}$. 
Therefore, the set of representations $\bm{6}^* \oplus \bm{6}'^* \oplus \bm{20}$ 
with appropriate $ ( \eta_{\mathcal{R}}, \, \eta'_{\mathcal{R}} )$ contain all quarks and leptons 
 of one generation and have no exotic fermions \cite{LM}.
In this way, we find that the set of representations containing one generation of quarks and leptons 
 without exotic fermions are only three types:
\begin{equation*}
	\begin{aligned}
		\bm{6}^* \oplus \bm{6'}^* \oplus \bm{20} &\qq{(Type 1)} \\
		\bm{15} \oplus \bm{15'} \oplus \bm{1} &\qq{(Type 2)} \\
		\bm{15} \oplus \bm{21} \oplus \bm{1} &\qq{(Type 3)}
	\end{aligned}
\end{equation*}
where an $SU(6)$ singlet $\bm{1}$ can incorporate a right-handed neutrino $\nu_R$. 
We would like to emphasize that new sets of representations with no exotic fermions are Type 2 and Type 3. 
These are found by allowing the $SU(6)$ singlets in a matter content. 
As a result, we obtain the models satisfying the condition ($\star$) by combining the above three types of representations,
for example, $ 2 \times (\bm{6}^* \oplus \bm{6}'^* \oplus \bm{20} ) + (\bm{15} \oplus \bm{21} \oplus \bm{1})$.
We note that $3 \times (\bm{6}^* \oplus \bm{6}'^* \oplus \bm{20})$ model was investigated in \cite{LM}.

A comment on the differences between \cite{ABBGW} and our model. 
In \cite{ABBGW}, the authors proposed that the quarks and leptons without exotic fermions per generation 
into the representations $\bm{20} \oplus \bm{15} \oplus \bm{6} \oplus \bm{1}$, 
which is obviously different from any sets of the representations we found. 
Furthermore, the representation $\bm{21}$ is also a possible choice in our analysis. 

\section{Higgs potential and Higgs mass}
We have obtained three types of representations in the previous section.
In this section, we will calculate one-loop Higgs potential and Higgs mass.
As we will see below, we introduce several fermions in the adjoint representation of $SU(6)$ 
to obtain correct electroweak symmetry breaking pattern $ SU(2)_L \times U(1)_Y \to U(1)_{em}$ 
 and a viable Higgs mass.
All matter fields we consider are adjoint fermions and 
the sets of representations including three generations of quarks and leptons
\begin{equation*}
	\begin{aligned}
		N_{\mathrm{ad}} \times \bm{35}
			&\oplus k_1 \times (\bm{6}^* \oplus \bm{6}'^* \oplus \bm{20}) \\
			&\oplus k_2 \times (\bm{15} \oplus \bm{15}' \oplus \bm{1}) \\
			&\oplus k_3 \times (\bm{15} \oplus \bm{21} \oplus \bm{1}) \, ,
	\end{aligned}
\end{equation*}
where $N_{\mathrm{ad}}$ is a number of adjoint fermions\footnote{In general, the exotic fermions 
appear from these adjoint fermions.
They are assumed to couple to the 4D brane localized fermions through Dirac mass terms.}
and $k_i \, ( i = 1, \, 2, \, 3)$ are non-negative integers satisfying $ k_1 + k_2 + k_3 = 3 $, 
 which implies three generations of quarks and leptons.
From (\ref{representations}), one can calculate the contributions of each representation 
to one-loop Higgs potential as follows. 
\begin{equation}
\label{eachV}
	\begin{aligned}
		V_{\mathrm{ad}} (\alpha)  &= \, C \Big[ 4N_{\mathrm{ad}} \sum_{ n = 1 }^{ \infty } 
		\frac{1}{n^5} \big\{ \cos{ ( 2 \pi n \alpha ) } + 2 \cos{ ( \pi n \alpha ) } + 6 (-1)^n \cos{ ( \pi n \alpha ) } \big\} \Big] \, ,\\
		V_g (\alpha) &= \, C \Big[ -3 \sum_{ n = 1 }^{ \infty } \frac{1}{n^5} \big\{ \cos{ ( 2 \pi n \alpha ) } + 2 \cos{ ( \pi n \alpha ) } 
		+ 6 (-1)^n \cos{ ( \pi n \alpha ) } \big\} \Big] \, ,\\
		V_1 (\alpha) &= \, C \Big[ 16 k_1 \sum_{ n = 1 }^{ \infty } \frac{ 1 + (-1)^n }{ n^5 } \cos{ ( \pi n \alpha ) } \Big] \, ,\\
		V_2 (\alpha) &= \, C \Big[ 16 k_2 \sum_{ n = 1 }^{ \infty } \frac{ 1 + (-1)^n }{ n^5 } \cos{ ( \pi n \alpha ) } \Big] \, ,\\
		V_3 (\alpha) &= \, C \Big[ 16 k_3 \sum_{ n = 1 }^{ \infty } \frac{ 1 + (-1)^n }{ n^5 } \cos{ ( \pi n \alpha ) } 
		+ 4 k_3 \sum_{ n = 1 }^{ \infty } \frac{ (-1)^n }{ n^5 } \cos{ ( 2 \pi n \alpha ) } \Big]\, ,
	\end{aligned}
\end{equation}
where $C$ is a constant as $C = 3/(128 \pi^7R^5)$. 
{$V_{\mathrm{ad}} (\alpha)$, $V_g (\alpha) $ and $V_i (\alpha) $ are the Higgs potential
by the contributions of adjoint fermions, gauge fields and fermions in Type $ i $, respectively.
$\alpha$ is a dimensionless parameter defined by
\begin{equation}
	 \expval{A_5} = \frac{\alpha}{2gR} \: \lambda_{27} \, ,
\end{equation}
where $\lambda_{27}$ is one of the $SU(6)$ generators, 
which has a value in the $(2, 6)$ component of $6 \times 6$ matrix.
In obtaining $V_2(\alpha)$ from $\bm{15}$ and $\bm{15'}$, $Z_2$ parity is taken as 
$( \eta_{\bm{15}}, \, \eta'_{\bm{15}} ) = (+, +)$ for $u_R, l_L, e_R$ 
and $( \eta_{\bm{15}}, \, \eta'_{\bm{15}} ) = (+, -)$ for $q_L, d_R$. 
From (\ref{eachV}), the total one-loop Higgs potential $V(\alpha)$ is given by
\begin{equation}
	\begin{aligned}
		V(\alpha) = \,
			&C \Big[ ( 4N_{\mathrm{ad}} - 3 ) \sum_{ n = 1 }^{ \infty } \frac{1}{n^5} \big\{ \cos{ ( 2 \pi n \alpha ) } + 2 \cos{ ( \pi n \alpha ) } + 6 (-1)^n \cos{ ( \pi n \alpha ) } \big\} \\
			&\quad + 48 \sum_{ n = 1 }^{ \infty } \frac{ 1 + (-1)^n }{ n^5 } \cos{ ( \pi n \alpha ) } + 4 k_3 \sum_{ n = 1 }^{ \infty } \frac{ (-1)^n }{ n^5 } \cos{ ( 2 \pi n \alpha ) }\Big] \, ,
	\end{aligned}
\end{equation}
where we used $ k_1 + k_2 + k_3 = 3 $.
For $ k_3 = 0 $, $V(\alpha) $ is same as the potential in $3 \times (\bm{6}^* \oplus \bm{6}'^* \oplus \bm{20})$ model 
and was discussed in \cite{LM}\footnote{Obviously, the other types of potentials with $k_3=0$, which are 
$2 \times (\mathrm{Type 1}) + 1 \times (\mathrm{Type 2}), \,
1 \times (\mathrm{Type 1}) + 2 \times (\mathrm{Type 2})$ and $3 \times (\mathrm{Type 2})$,  
are reduce to that of $3 \times (\mathrm{Type 1}) $.}.
Here, we will analyze one-loop Higgs potential in the cases of $k_3 = 1, \;  2, \;  3$.
\begin{figure}[h]
	\begin{minipage}[b]{0.3\linewidth}
		\centering
		\includegraphics[width = 0.8\linewidth]{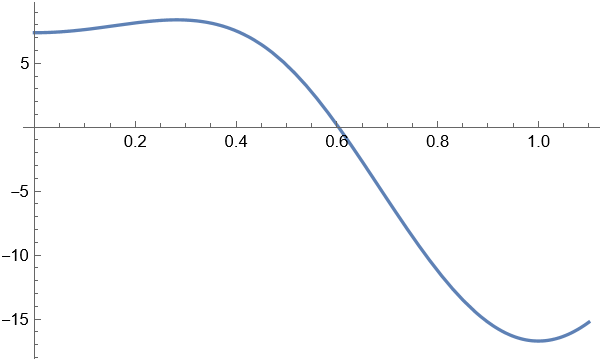}
		\subcaption{$k_3 = 1, \; N_{\mathrm{ad}}= 0$}
	\end{minipage}
	\begin{minipage}[b]{0.3\linewidth}
		\centering
		\includegraphics[width = 0.8\linewidth]{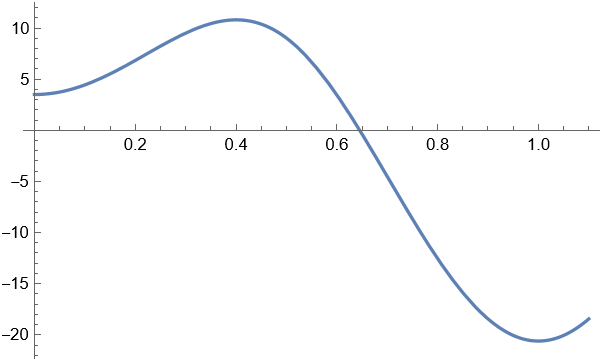}
		\subcaption{$k_3 = 2, \;N_{\mathrm{ad}}= 0$}
	\end{minipage}
	\begin{minipage}[b]{0.3\linewidth}
		\centering
		\includegraphics[width = 0.8\linewidth]{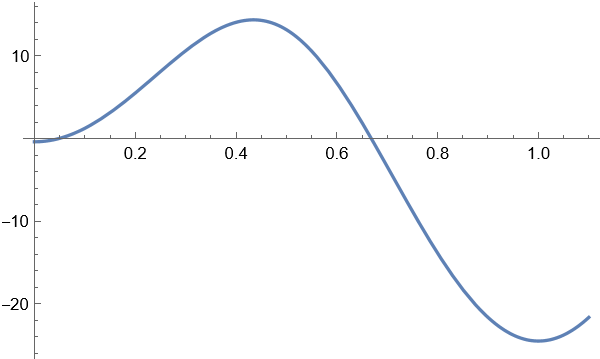}
		\subcaption{$k_3 = 3, \; N_{\mathrm{ad}}= 0$}
	\end{minipage}
	\caption{For $k_3 = 1, \, 2, \, 3$, the plots of one-loop Higgs potential $V(\alpha)/C$ with no adjoint fermions are shown.}
\end{figure}
\begin{figure}[h]
	\begin{minipage}[b]{0.3\linewidth}
		\centering
		\includegraphics[width = 0.8\linewidth]{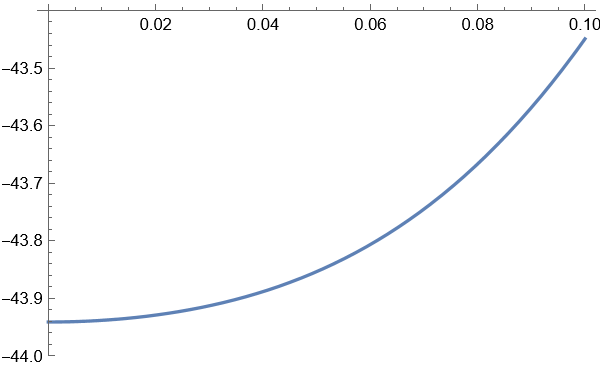}
		\subcaption{$k_3 = 3, \; N_{\mathrm{ad}}= 4$}
	\end{minipage}
	\begin{minipage}[b]{0.3\linewidth}
		\centering
		\includegraphics[width = 0.8\linewidth]{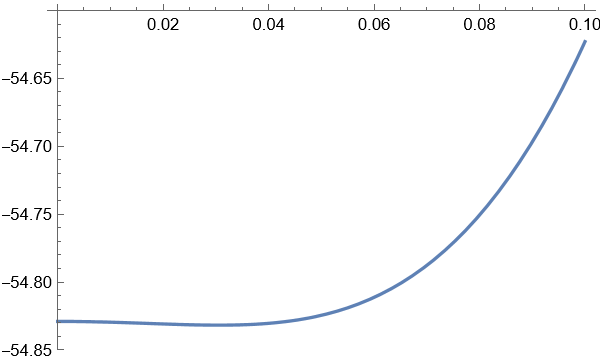}
		\subcaption{$k_3 = 3, \;N_{\mathrm{ad}}= 5$}
	\end{minipage}
	\begin{minipage}[b]{0.3\linewidth}
		\centering
		\includegraphics[width = 0.8\linewidth]{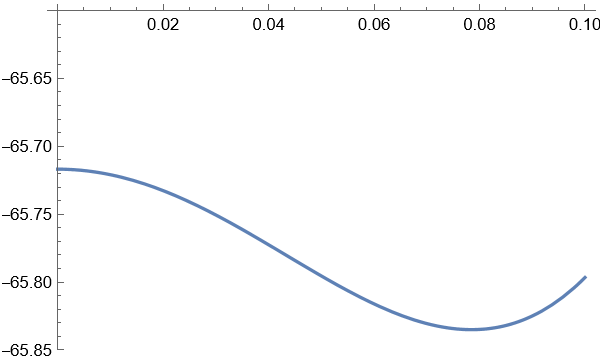}
		\subcaption{$k_3 = 3, \; N_{\mathrm{ad}}= 6$}
	\end{minipage}
	\caption{For $k_3 = 3$, the plots of one-loop Higgs potential $V(\alpha)/C$ 
	with four, five and six adjoint fermions are shown.}
\end{figure}
\begin{table}[h]
	\centering
	\begin{tabular}{c|c|c|c} \hline
		{} & $N_{\mathrm{ad}}$ & $\alpha_0$ & $ SU(2)_L \times U(1)_Y \to U(1)_{em}$ \\ \hline \hline
		{} & 1 & 0.104072 & $\checkmark$ \\
		$k_3 = 1$ & 2 & 0.171499 & $\checkmark$ \\
		{} & $\vdots$ & $\vdots$ & $\vdots$ \\ \hline
		{} & 2 & 0 & $\times$ \\
		$k_3 = 2$ & 3 & 0.0443208 & $\checkmark$ \\
		{} & 4 & 0.103231 & $\checkmark$ \\
		{} & $\vdots$ & $\vdots$ & $\vdots$ \\ \hline
		{} & 4 & 0 & $\times$ \\
		$k_3 = 3$ & 5 & 0.0305824 & $\checkmark$ \\
		{} & 6 & 0.0785057 & $\checkmark$ \\
		{} & $\vdots$ & $\vdots$ & $\vdots$ \\ \hline
	\end{tabular}
	\caption{For possible values of $k_3$, the required number of adjoint fermions $N_{\mathrm{ad}}$
			to obtain the electroweak symmetry breaking $SU(2)_L \times U(1)_Y \to U(1)_{em}$ 
			and corresponding VEV of Higgs field $\alpha_0$ are shown. 
			$\checkmark$ means viable solutions.}
\end{table} \\
As can be seen from Figure 1, 
 we cannot obtain a desired electroweak symmetry breaking, 
 if adjoint fermions are not introduced.  
This is because the vacuum expectation value (VEV) of the potential minimum $\alpha_0$ is unity, $\alpha_0 = 1$, 
 an extra $U(1)$ other than $U(1)_{em}$ gauge symmetry remains unbroken in this case. 
With several adjoint fermions,  
$\alpha_0 $ is located in a range $0 < \alpha_0 < 1$ and the desired electroweak symmetry breaking is realized. 
Some results are shown in Table 2.
In fact, we need more adjoint fermions to obtain a realistic Higgs mass.
Higgs mass $m_H$ can be calculate from $\mathrm{d}^2V/\mathrm{d}\alpha^2$ as
\begin{equation}
	\begin{aligned}
		m_H^2 
		= \, & g^2 R^2 \eval{ \dv[2]{V}{\alpha}}_{\alpha = \alpha_0} \\
		= \; & - \frac{3 g_4^2 M_W^2}{64 \pi ^4 \alpha_0^2}
				\Big[ ( 4N_{\mathrm{ad}} - 3 ) 
				\sum_{ n = 1 }^{ \infty } \frac{1}{n^3} \big\{ 4 \cos{ ( 2 \pi n \alpha_0 ) } + 2 \cos{ ( \pi n \alpha_0 ) } 
				+ 6 (-1)^n \cos{ ( \pi n \alpha_0 ) } \big\} \\
			&\hspace{5em} + 48 \sum_{ n = 1 }^{ \infty } \frac{ 1 + (-1)^n }{ n^3 } \cos{ ( \pi n \alpha_0 ) } 
			+ 16 k_3 \sum_{ n = 1 }^{ \infty } \frac{ (-1)^n }{ n^3 } \cos{ ( 2 \pi n \alpha_0 ) }\Big], 
	\end{aligned}	
\end{equation}
where we used the fact that
$W$ boson mass $M_W$ is given by $M_W = \alpha_0 / R $.
Note also that the gauge coupling constants $g_4$ in 4D
and $g$ in 5D are related by $ g_4^2 = g^2/(2 \pi R ) $.
We illustrate that Higgs mass $125$ GeV can be indeed obtained and our results are shown in Table 3.
In this calculation, the 4D gauge coupling constant $g_4$ is taken to be an experimental value around 0.7\footnote{
More precisely, 4D gauge coupling constant $g_4$ in Higgs potential $V(\alpha)$ should be evaluated 
at the compactification scale not the weak scale.
In our model, the compactification scale is considered to be around 10 TeV 
and the renormalization effects between these two scales are expected to be small.
Therefore, we employed $ g_4 \sim 0.7 $ in evaluating the Higgs mass.}.
\begin{table}[h]
	\centering
	\begin{tabular}{c|c|c|c} \hline
		{} & $N_{\mathrm{ad}}$ & $\alpha_0$ & Higgs mass \\ \hline \hline
		$k_3 = 0$ & 47 & 0.210822 & 177.002 $g_4$ GeV \\
		{} 		& 48 & 0.210736 & 178.913 $g_4$ GeV \\ \hline
		$k_3 = 1$ & 46 & 0.205474 & 176.288 $g_4$ GeV \\
		{} 		& 47 & 0.205498 & 178.208 $g_4$ GeV \\ \hline
		$k_3 = 2$ & 46 & 0.199885 & 177.641 $g_4$ GeV \\
		{}		& 47 & 0.200031 & 179.641 $g_4$ GeV \\ \hline
		$k_3 = 3$ & 45 & 0.193864 & 177.238 $g_4$ GeV \\
		{}		& 46 & 0.194151 & 179.138 $g_4$ GeV \\ \hline
	\end{tabular}
	\caption{For possible values of $k_3$, the required number of adjoint fermions $N_{\mathrm{ad}}$ 
	and the VEV of Higgs fields $\alpha_0$ to obtain the experimental value of Higgs mass 125 GeV are shown.
			4D gauge coupling constant $g_4$ is taken to be an experimental value $g_4 \sim 0.7$.}
\end{table} \\
The results for Higgs potential analysis are given just for illustration.
To study in a more realistic setup, we have to consider more improvements. \\
\indent
In our analysis, all of the fermions are assumed to be massless.
One of the ways to incorporate the SM fermion masses is introducing a $Z_2$ odd bulk mass $M \epsilon (y) $ 
for 5D fermion masses where $ \epsilon(y) $ is a sign function. 
Then, the quark and lepton masses can be obtained by an overlap integral of zero mode functions \cite{AHS}.
Then, 
$V_i(\alpha)$ are modified as 
\begin{equation}
\label{modifiedV}
	\begin{aligned}
V_1 (\alpha) &= \, C \Big[ 16 k_1 \sum_{ n = 1 }^{ \infty } \frac{ 1 + (-1)^n }{ n^5 }  f(2 \pi R M_1) \cos{ ( \pi n \alpha ) } \Big] \, ,\\
V_2 (\alpha) &= \, C \Big[ 16 k_2 \sum_{ n = 1 }^{ \infty } \frac{ 1 + (-1)^n }{ n^5 } f(2 \pi R M_2) \cos{ ( \pi n \alpha ) } \Big] \, ,\\
V_3 (\alpha) &= \, C \Big[ 16 k_3 \sum_{ n = 1 }^{ \infty } \frac{ 1 + (-1)^n }{ n^5 }  f( 2 \pi R M_3)  \cos{ ( \pi n \alpha ) } 
+ 4 k_3 \sum_{ n = 1 }^{ \infty } \frac{ (-1)^n }{ n^5 }  f( 2 \pi R M_3)  \cos{ ( 2 \pi n \alpha ) }  \Big] \, ,
	\end{aligned}
\end{equation}
where 
$ M_i $ are bulk mass of 
fermions in $i$th generation respectively, 
and $f(z)$ is defined as
\begin{equation}
	f(z) = \qty( 1 + nz + \frac{1}{3} n^2 z^2) e^{-nz}.
\end{equation}
The modification to obtain (\ref{modifiedV}) is merely an insertion of  $f( 2 \pi R M )$.
From this modification, we need only a little change of the value of $N_{\mathrm{ad}}$ 
for the desired symmetry breaking and Higgs mass,
and remarkable results are not expected. 
Therefore, we do not perform this analysis further.

\section{Conclusion}
In this paper, we have discussed a 5D $SU(6)$ GGHU, 
where we discovered new sets of fermion matter content 
including only quarks and leptons per generation as massless fields. 
After a systematic group theoretical search is performed, 
we have found three types of $SU(6)$ representation sets providing only quarks and leptons as massless fermions, 
$\bm{6}^* \oplus \bm{6'}^* \oplus \bm{20}$ (Type 1), 
$\bm{15} \oplus \bm{15}' \oplus \bm{1}$ (Type 2), 
$\bm{15} \oplus \bm{21} \oplus \bm{1}$ (Type 3). 
Type 1 has been already discussed in \cite{LM}, but Type 2 and Type 3 are newly discovered 
by allowing a $SU(6)$ gauge singlet for the right-handed neutrino. 
Comparing to Type 1 which has a three-rank tensor representation, 
it is interesting to see that Type 2 and Type 3 have at most two-rank tensor representations.  

One-loop Higgs potential was calculated for the fermions with a combination of various types 
of the above mentioned representation sets and extra adjoint fermions in addition to the gauge fields. 
Without extra adjoint fermions, the electroweak symmetry breaking cannot be realized. 
Introducing some extra adjoint fermions has led to the correct electroweak symmetry breaking and a viable Higgs mass. 
We have explicitly shown some numerical examples as an illustration. 
In the Higgs potential analysis performed in this paper, 
a viable Higgs mass is obtained by introducing a relatively many adjoint fermions. 
Simplifying the fermion representations is a challenging problem. 

In order to construct a more realistic model, there are many issues to be explored. 

In GGHU scenario, the construction of Yukawa coupling is a very nontrivial issue. 
Naively, Yukawa coupling can be generated from the gauge coupling in the 5D bulk 
if the left-handed and the right-handed quark(lepton) belong to the same multiplet. 
The up-type quark and the neutrino Yukawa couplings in Type 1 \cite{LM} and  
the down-type quark and the charged lepton Yukawa couplings in Type 2, 3 can be generated from the gauge coupling. 
As for the remaining Yukawa couplings, 
it has to manage to construct non-local Yukawa couplings 
by an interplay between the bulk and the brane localized fields \cite{SSS, MYT}. 
For Type 2, if the following interaction is allowed in the 5D gauge interaction, 
the up-type quark Yukawa coupling can be obtained 
from the mixing terms between $\bm{15}$ and $\bm{15'}$. 
\begin{align}
(\overline{\bm{15} + c(y) \bm{15'}}) \gamma^5 A_5 (\bm{15} + c(y) \bm{15'}),
\end{align}
where the parities of $\bm{15}, \bm{15'}$ and some function $c(y)$ are $(+, +), (+,-)$ 
and $(+,-)$, respectively, for example. 
If this is possible, all Yukawa couplings can be generated from the gauge coupling in the 5D bulk. 
Then, whether a realistic fermion mass hierarchy, generation mixing, CP phase 
can be obtained is a next problem to be solved. 

The gauge coupling unification and the proton decay analysis are also important. 
In \cite{MYT}, the gauge coupling unification in $SU(6)$ GGHU was studied. 
Interestingly, the running is almost logarithmic in the energy scale 
and the unification scale is relatively high $\sim 10^{14}$ GeV, which is different from \cite{DDG}. 
This is because the multiplets in the 5D bulk respect $SU(6)$ symmetry in our GGHU, 
but those in \cite{DDG} respect only the SM gauge group. 
This feature would be also expected in our new GGHU case. 
As for the proton decay analysis, 
the high scale of unification is a good point as mentioned above, 
but the X, Y gauge boson mass is very small, which is around the compactification scale $\sim$ 10 TeV. 
Therefore, the proton rapidly decays and our model is immediately excluded by SuperKamiokande data. 
In order to avoid this situation, the dangerous baryon number violating interactions have to be forbidden 
by some symmetry, for instance, by an extra $U(1)$ symmetry or its unbroken discrete symmetry. 
It would be also interesting to study the main mode of the proton decay in our model.  

These issues are left for our future work. 



\appendix
\section{Classification of boundary conditions}
In this section, we study the diagonal parity matrices $\hat{P}$ and $\hat{P'}$ 
by which the symmetry breaking $SU(6) \to SU(3)_c \times SU(2)_L \times U(1)_Y \times U(1)_X$ is realized
and we can obtain a massless Higgs doublet and make a colored Higgs field massive.
As shown in \cite{HHK}, we can always diagonalize $\hat{P}$ and $\hat{P'}$, 
which are specified by three non-negative integers $(p, q, r)$ such as
\begin{equation}
	\begin{aligned}
		P = \; & \mathrm{diag} \overbrace{(+1, \dots, +1, \; +1, \dots, +1, \; -1, \dots, -1,  \; -1, \dots, -1 )}^{N}, \\
		P' = \; & \mathrm{diag}\underbrace{( +1, \dots, +1}_{p}, \; \underbrace{-1, \dots, -1}_{q}, \; 
		\underbrace{+1, \dots, +1}_{r}, \; \underbrace{-1, \dots, -1)}_{s = N - p - q - r}
	\end{aligned}
\end{equation}
where $N \geq p + q + r$.
These parities break $SU(N)$ to $SU(p) \times SU(q) \times SU(r) \times SU(s) \times U(1)^4$.
We write the set of these parities as $[p;q,r;s]$.
Since we would like to consider the symmetry breaking $SU(6) \to SU(3) \times SU(2) \times U(1)^2$,
we take $N = 6$ and the set $(p, q, r, s)$ corresponds to the permutation of $(0,1,2,3)$.
We next consider which kind of $[p;q,r;s]$ should be chosen in order not to obtain a massless colored Higgs field.
For parities $[p;q,r;s]$, $Z_2$ parity assignments of matrix elements of $A_5$ is given by
\begin{equation*}
	A_{5} = 
		\begin{aligned}
		&\hspace{0.7em}\overbrace{\hspace{9.7em}}^p \overbrace{\hspace{9.4em}}^q \overbrace{\hspace{9.4em}}^r \overbrace{\hspace{9.7em}}^s \\[-2mm]
		&\left(
		\begin{array}{ccc|ccc|ccc|ccc}
			(-,-) & \cdots & (-,-) & (-,+) & \cdots & (-,+) & (+,-) & \cdots & (+,-) & (+,+) & \cdots & (+,+)\\
			\vdots & {} & \vdots & \vdots & {} & \vdots & \vdots & {} & \vdots & \vdots & {} & \vdots \\
			(-,-) & \cdots & (-,-) & (-,+) & \cdots & (-,+) & (+,-) & \cdots & (+,-) & (+,+) & \cdots & (+,+)\\[2pt] 
			\hline \vspace{-10pt}
			{} & {} & {} & {} & {} & {} & {} & {} & {} & {} & {} & {} \\
			(-,+) & \cdots & (-,+) & (-,-) & \cdots & (-,-) & (+,+) & \cdots & (+,+) & (+,-) & \cdots &  (+,-)\\
			\vdots & {} & \vdots & \vdots & {} & \vdots & \vdots & {} & \vdots & \vdots & {} & \vdots \\
			(-,+) & \cdots & (-,+) & (-,-) & \cdots & (-,-) & (+,+) & \cdots & (+,+) & (+,-) & \cdots & (+,-)\\[2pt] 
			\hline \vspace{-10pt}
			{} & {} & {} & {} & {} & {} & {} & {} & {} & {} & {} & {} \\
			(+,-) & \cdots & (+,-) & (+,+) & \cdots & (+,+) & (-,-) & \cdots & (-,-) & (-,+) & \cdots & (-,+)\\
			\vdots & {} & \vdots & \vdots & {} & \vdots & \vdots & {} & \vdots & \vdots & {} & \vdots \\
			(+,-) & \cdots & (+,-) & (+,+) & \cdots & (+,+) & (-,-) & \cdots & (-,-) & (-,+) & \cdots & (-,+)\\[2pt] 
			\hline \vspace{-10pt}
			{} & {} & {} & {} & {} & {} & {} & {} & {} & {} & {} & {} \\
			(+,+) & \cdots & (+,+) & (+,-) & \cdots & (+,-) & (-,+) & \cdots & (-,+) & (-,-) & \cdots & (-,-)\\
			\vdots & {} & \vdots & \vdots & {} & \vdots & \vdots & {} & \vdots & \vdots & {} & \vdots \\
			(+,+) & \cdots & (+,+) & (+,-) & \cdots & (+,-) & (-,+) & \cdots & (-,+) & (-,-) & \cdots & (-,-)
		\end{array}
		\right) \, .
		\end{aligned}
\end{equation*}
We find that the $A_5$ has massless zero modes in blocks 
as $(\bm{p}, \bm{1}, \bm{1}, \bm{s})$ and $(\bm{1}, \bm{q}, \bm{r}, \bm{1})$ 
which expresses the representations of ($SU(p), SU(q), SU(r), SU(s))$.
There are only eight choices of $[p;q,r;s]$ to obtain a massless Higgs doublet and make a colored Higgs field massive. 
\begin{equation}
	\begin{aligned}
	(a) \; [3;2,1;0], \qquad &(b) \; [1;0,3;2], \\
	(c) \; [0;1,2;3], \qquad &(d) \; [2;3,0;1], \\
	(e) \; [2;0,3;1], \qquad &(f) \; [3;1,2;0], \\
	(g) \; [1;3,0;2], \qquad &(h) \; [0;2,1;3].
	\end{aligned}
\end{equation}
As we will show below, these parities are found to be equivalent.
One can easily find that physical differences does not appear by interchanging $\hat{P} \leftrightarrow \hat{P'}$.
If we set $-\hat{P}$ instead of $\hat{P}$, (\ref{fermiontr.}) is rewritten as
\begin{equation}
		\psi_{\mathcal{R}} \to \eta_{\mathcal{R}} \hat{P}(\mathcal{R}) \gamma^5 \psi_{\mathcal{R}}\qq{at} y = 0.
\end{equation}
Redefining $\eta_{\mathcal{R}}$ as $-\eta_{\mathcal{R}}$, the above formula is equivalent to (\ref{fermiontr.}).
Therefore, $\hat{P}$ and $-\hat{P}$ are physically equivalent.
The same argument holds for $\hat{P'}$.
Interchanging $\hat{P} \leftrightarrow \hat{P'}$ is expressed by interchanging $q \leftrightarrow r$, that is, 
\begin{equation*}
	[p;q.r;s] \sim [p;r,q;s],
\end{equation*}
where the symbol $\sim$ means the equivalence between the two sets of the parities .
Therefore, we find
\begin{equation}
\label{d sim e}
	(d) \; [0;1,2;3] \sim (e) \; [0;2,1;3] \, .
\end{equation}
Next, changing the sign $\hat{P} \to -\hat{P}$ and $\hat{P'} \to -\hat{P'}$ are expressed 
by interchanging $(p,q) \leftrightarrow (r,s)$ and $(p,r) \leftrightarrow (q,s)$ respectively, that is,
\begin{equation*}
	\begin{aligned}
		&[p;q.r;s] \sim [r;s,p;q] \, , \\
		&[p;q,r;s] \sim [q;p,s;r] \, .
	\end{aligned}
\end{equation*}
Thus, we find the following equivalence relations. 
\begin{equation}
	\begin{aligned}
		&(a) \; [3;2,1;0] \sim (b) \; [1;0,3;2] \sim (c) \; [0;1,2;3] \sim (d) \; [2;3,0;1]\, , \\
		&(e) \; [2;0,3;1] \sim (f) \; [3;1,2;0] \sim (g) \; [1;3,0;2] \sim (h) \; [0;2,1;3]\, .
	\end{aligned}
\end{equation}
Combining (\ref{d sim e}), all of the sets of $Z_2$ parity are equivalent. 
\begin{equation}
	(a) \sim (b) \sim (c) \sim (d) \sim (e) \sim (f) \sim (g) \sim (h).
\end{equation}
Therefore, we conclude that it is sufficient to consider the case $(a)$.

\section{Systematic search for the representations without exotic fermions}
In this section, we study that what kind of representations include exotic fermions 
and we finally find the representations without exotic fermions.
General irreducible representation of $SU(6)$ $\mathcal{R}$ corresponds to the following Young tableau 
\begin{equation*}
	\ytableausetup{ boxsize = normal }
	\begin{aligned}
		&\overbrace{\hspace{4.45em}}^j \overbrace{\hspace{4.45em}}^k \overbrace{\hspace{4.45em}}^l 
		\overbrace{\hspace{4.45em}}^m \overbrace{\hspace{4.45em}}^n \\[-2mm]
		&\begin{ytableau}
			{} & \cdots & {} & {} & \cdots & {} & {} & \cdots & {} & {} & \cdots & {} & {} & \cdots & {} \\
			{} & \cdots & {} & {} & \cdots & {} & {} & \cdots & {} & {} & \cdots & {} \\
			{} & \cdots & {} & {} & \cdots & {} & {} & \cdots & {} \\
			{} & \cdots & {} & {} & \cdots & {} \\
			{} & \cdots & {} \\
		\end{ytableau}
	\end{aligned}
\end{equation*}
where $j, \; k, \; l, \; m$ and $n$ are non-negative integers.
Let us discuss in order if exotic fermions exist or not.
\begin{enumerate}
	\item $ n \geq 1 $ . \\
		Let us rewrite $\mathcal{R}$ as
		\begin{equation*}
			\mathcal{R} = \;
				\begin{aligned}
					&\overbrace{\hspace{4.45em}}^j \overbrace{\hspace{4.45em}}^k \overbrace{\hspace{4.45em}}^l \overbrace{\hspace{4.45em}}^m \overbrace{\hspace{6em}}^n \\[-2mm]
					&\begin{ytableau}
						{} & \cdots & {} & {} & \cdots & {} & {} & \cdots & {} & {} & \cdots & {} & {} & \cdots & {} & a \\
						{} & \cdots & {} & {} & \cdots & {} & {} & \cdots & {} & {} & \cdots & {} \\
						{} & \cdots & {} & {} & \cdots & {} & {} & \cdots & {} \\
						{} & \cdots & {} & {} & \cdots & {} \\
						{} & \cdots & {} \\
					\end{ytableau}
				\end{aligned}
				\, .
		\end{equation*}
		Decomposing $\mathcal{R}$ into representations of $SU(3)_c \times SU(2)_L \times U(1)_Y \times U(1)_X$ and 
		focusing on a following blocks as
		\begin{equation*}
			\begin{aligned}
				&\begin{ytableau}
					{} & \cdots & {} & {} & \cdots & {} & {} & \cdots & {} & {} & \cdots & {} & {} & \cdots & {} & \none & a \\
					{} & \cdots & {} & {} & \cdots & {} & {} & \cdots & {} & {} & \cdots & {} \\
					\none \\
					{} & \cdots & {} & {} & \cdots & {} & {} & \cdots & {} \\
					{} & \cdots & {} & {} & \cdots & {} \\
					\none \\
					{} & \cdots & {} \\
				\end{ytableau}
			\end{aligned}
			\, ,
		\end{equation*}
		one finds that $\mathcal{R}$ contains fields $ \psi_1, \, \psi_2 $ with following representations
		\begin{equation*}
			\begin{aligned}
			\ytableausetup{ boxsize = 1.3em }
				&\psi_1 : \qty( \; 
					\raisebox{3mm}{ {\footnotesize$
					\begin{aligned}
						&\overbrace{\hspace{3.85em}}^j \overbrace{\hspace{3.85em}}^k 
						\overbrace{\hspace{3.85em}}^l \overbrace{\hspace{3.85em}}^m 
						\overbrace{\hspace{5.2em}}^n
						\\[-2mm]
						&\begin{ytableau}
							{} & \cdots & {} & {} & \cdots & {} & {} & \cdots & {} & {} & \cdots & {} & {} & \cdots & {} & a \\
							{} & \cdots & {} & {} & \cdots & {} & {} & \cdots & {} & {} & \cdots & {} 
						\end{ytableau}
					\end{aligned} $}} \raisebox{-4mm}{,} \: 
					\raisebox{3mm}{ {\footnotesize$
					\begin{aligned}
						&\overbrace{\hspace{3.8em}}^j \overbrace{\hspace{3.8em}}^k 
						\overbrace{\hspace{3.8em}}^l \\[-2mm]
						&\begin{ytableau}
							{} & \cdots & {} & {} & \cdots & {} & {} & \cdots & {} \\
							{} & \cdots & {} & {} & \cdots & {} 
						\end{ytableau}
					\end{aligned} $}} \raisebox{-4mm}{,} \: 
					\raisebox{3mm}{ {\footnotesize$
					\begin{aligned}
						&\overbrace{\hspace{3.8em}}^j \\[-2mm]
						&\begin{ytableau}
							{} & \cdots & {} 
						\end{ytableau}
					\end{aligned} \; $}} ) \, ,\\
				&\psi_2 : \qty( \; 
					\raisebox{3mm}{ {\footnotesize$
					\begin{aligned}
						&\overbrace{\hspace{3.85em}}^j \overbrace{\hspace{3.85em}}^k 
						\overbrace{\hspace{3.85em}}^l \overbrace{\hspace{3.85em}}^m 
						\overbrace{\hspace{3.85em}}^{n - 1}
						\\[-2mm]
						&\begin{ytableau}
							{} & \cdots & {} & {} & \cdots & {} & {} & \cdots & {} & {} & \cdots & {} & {} & \cdots & {} \\
							{} & \cdots & {} & {} & \cdots & {} & {} & \cdots & {} & {} & \cdots & {} 
						\end{ytableau}
					\end{aligned} $}} \raisebox{-4mm}{,} \: 
					\raisebox{3mm}{ {\footnotesize$
					\begin{aligned}
						&\overbrace{\hspace{3.85em}}^j \overbrace{\hspace{3.85em}}^k 
						\overbrace{\hspace{5.2em}}^{l + 1}\\[-2mm]
						&\begin{ytableau}
							{} & \cdots & {} & {} & \cdots & {} & {} & \cdots & {} & a \\
							{} & \cdots & {} & {} & \cdots & {} 
						\end{ytableau}
					\end{aligned} $}} \raisebox{-4mm}{,} \: 
					\raisebox{3mm}{ {\footnotesize$
					\begin{aligned}
						&\overbrace{\hspace{3.85em}}^j \\[-2mm]
						&\begin{ytableau}
							{} & \cdots & {} 
						\end{ytableau}
					\end{aligned} \; $}} ) \, .\\
				\end{aligned}
			\end{equation*}
			where boxes in the parenthesis mean the same as (\ref{56}) in Section 3.
			$Z_2$ parities $(P, P')$ of ${\psi_1}_L \, , \, {\psi_1}_R \, ,\, {\psi_2}_L $ and ${\psi_2}_R$ are given as
			\begin{equation}
				\begin{alignedat}{3}
					{\psi_1}_L :\, &\big( &   &(-1)^j \eta_{\mathcal{R}}, &	  &(-1)^l \eta_{\mathcal{R}}' \, \big) \, ,\\
					{\psi_1}_R :\, &\big( &  -&(-1)^j \eta_{\mathcal{R}}, & \; -&(-1)^l \eta_{\mathcal{R}}' \, \big) \, ,\\
					{\psi_2}_L :\, &\big( &   &(-1)^j \eta_{\mathcal{R}}, & 	 -&(-1)^l \eta_{\mathcal{R}}' \, \big) \, ,\\
					{\psi_2}_R :\, &\big( & -&(-1)^j \eta_{\mathcal{R}}, &  	  &(-1)^l \eta_{\mathcal{R}}' \, \big) \, .
				\end{alignedat}
			\end{equation}
			%
			Since we see all patterns of $(P, P')$, one of them has a massless zero mode.
			Let us find conditions consistent with the condition ($\star$).
			Focusing on $SU(3)$ and $SU(2)$ representations of $\psi_{1,2}$, 
			the following conditions are found to be satisfied.
			\begin{enumerate}
				\item If $ n \geq 3 $, $\mathcal{R}$ is not consistent with the condition $(\star)$.
				\item If $ n = 2 $, only when  $j = k = l = m = 0$, $\mathcal{R}$ is consistent with the condition ($\star$).
				\item If $ n = 1 $, only when  $ j + k + l + m \leq 1 $, $\mathcal{R}$ is consistent with the condition ($\star$).
			\end{enumerate}
		From this observation, it is sufficient to consider the representations below.
		\begin{equation*}
			\mathcal{R} =
				\bm{6}\, , \, \bm{21} \, , \, \bm{70} \, , \, \bm{84} \, , \, \bm{35}
				= 
				\ytableausetup{ boxsize = normal, centertableaux}
				\begin{ytableau}
					{} 
				\end{ytableau}
				\; , \;
				\begin{ytableau}
					{} & {} 
				\end{ytableau}
				\; , \,
				\begin{ytableau}
					{} & {} \\
					{} 
				\end{ytableau}
				\; , \;
				\begin{ytableau}
					{} & {} \\
					{} \\
					{} \\
					{} 
				\end{ytableau}
				, \;
				\begin{ytableau}
					{} & {} \\
					{} \\
					{} \\
					{} \\
					{}
				\end{ytableau} \, .
		\end{equation*}
		However, not all of them are available.
		In fact, $\bm{70}$ necessarily contains exotic fermions because it is  decomposed as
		\begin{equation*}
			\begin{aligned}
				\ytableausetup{ boxsize = 1em, centertableaux}
				\begin{ytableau}
					{} & a \\
					{} \\
				\end{ytableau} \; = \;
					&\qty( \;
						\begin{ytableau}
							{} & a  \\
							{} 
						\end{ytableau} \; , \quad
						\bullet \; , \quad
						\bullet \;
					)_L^{ ( \eta_{\bm{70}} , -\eta'_{\bm{70}} ) }
					\oplus
					\qty( \;
						\begin{ytableau}
							{} & a  \\
							{} 
						\end{ytableau} \; , \quad
						\bullet \; , \quad
						\bullet \;
					)_R^{ ( -\eta_{\bm{70}} , \eta'_{\bm{70}} ) } \\[3mm]
					& \qquad
					\oplus
					\Big( \;
						\begin{ytableau}
							{} & a \\
						\end{ytableau} \; , \quad
						\begin{ytableau}
							{} 
						\end{ytableau} \; , \quad
						\bullet \;
					\Big)_L^{ ( \eta_{\bm{70}} , \eta'_{\bm{70}} ) }
					\oplus
					\Big( \;
						\begin{ytableau}
							{} & a \\
						\end{ytableau} \; , \quad
						\begin{ytableau}
							{}
						\end{ytableau} \; , \quad
						\bullet \;
					\Big)_R^{ ( -\eta_{\bm{70}} , -\eta'_{\bm{70}} ) } 
					\oplus \cdots \; .
				\end{aligned}
		\end{equation*}
		In the same way, we find that $\bm{84}$ has exotic fermions, too:
		\begin{equation*}
			\begin{aligned}
				\begin{ytableau}
					{} & a \\
					{} \\
					{} \\
					{}
				\end{ytableau} = \;
					&\qty( \;
						\begin{ytableau}
							{} & a \\
							{} 
						\end{ytableau} \; , \quad
						\begin{ytableau}
							{} \\
							{}
						\end{ytableau} \; , \quad
						\bullet \;
					)_L^{ ( \eta_{\bm{84}} , \eta'_{\bm{84}} ) }
					\oplus
					\qty( \;
						\begin{ytableau}
							{} & a \\
							{} 
						\end{ytableau} \; , \quad
						\begin{ytableau}
							{} \\
							{}
						\end{ytableau} \; , \quad
						\bullet \;
					)_R^{ ( -\eta_{\bm{84}} , -\eta'_{\bm{84}} ) } \\
					& \qquad
					\oplus
					\qty( \;
						\begin{ytableau}
							{} & a \\
						\end{ytableau} \; , \quad
						\begin{ytableau}
							{} \\
							{}
						\end{ytableau} \; , \quad
						\begin{ytableau}
							{} 
						\end{ytableau} \;
					)_L^{ ( -\eta_{\bm{84}} , \eta'_{\bm{84}} ) }
					\oplus
					\qty( \;
						\begin{ytableau}
							{} & a \\
						\end{ytableau} \; , \quad
						\begin{ytableau}
							{} \\
							{}
						\end{ytableau} \; , \quad
						\begin{ytableau}
							{} 
						\end{ytableau} \;
					)_R^{ ( \eta_{\bm{84}} , -\eta'_{\bm{84}} ) } 
					\oplus \cdots \; .
				\end{aligned}
		\end{equation*}
		On the other hand, $\bm{35}$ is decomposed as 
		\begin{equation*}
			\begin{aligned}
				\begin{ytableau}
					{} & a \\
					{} \\
					{} \\
					{} \\
					{}
				\end{ytableau} = \;
					&\qty( \;
						\begin{ytableau}
							{} & a \\
							{} \\
							{}
						\end{ytableau} \; , \quad
						\begin{ytableau}
							{}
						\end{ytableau} \; , \quad
						\begin{ytableau}
							{} 
						\end{ytableau} \;
					)_L^{ ( -\eta_{\bm{35}} , -\eta'_{\bm{35}} ) }
					\oplus
					\qty( \;
						\begin{ytableau}
							{} & a \\
							{} \\
							{}
						\end{ytableau} \; , \quad
						\begin{ytableau}
							{}
						\end{ytableau} \; , \quad
						\begin{ytableau}
							{}
						\end{ytableau} \;
					)_R^{ ( \eta_{\bm{35}} , \eta'_{\bm{35}} ) } \\
					&\qquad
					\oplus
					\qty( \;
						\begin{ytableau}
							{} & a \\
							{}
						\end{ytableau} \; , \quad
						\begin{ytableau}
							{} \\
							{}
						\end{ytableau} \; , \quad
						\begin{ytableau}
							{} 
						\end{ytableau} \;
					)_L^{ ( -\eta_{\bm{35}} , \eta'_{\bm{35}} ) }
					\oplus
					\qty( \;
						\begin{ytableau}
							{} & a \\
							{}
						\end{ytableau} \; , \quad
						\begin{ytableau}
							{} \\
							{}
						\end{ytableau} \; , \quad
						\begin{ytableau}
							{}
						\end{ytableau} \;
					)_R^{ ( \eta_{\bm{35}} , -\eta'_{\bm{35}} ) } \\[2mm]
					& \qquad
					\oplus
					\qty( \;
						\begin{ytableau}
							{} \\
							{}
						\end{ytableau} \; , \quad
						\begin{ytableau}
							{} & a \\
							{}
						\end{ytableau} \; , \quad
						\begin{ytableau}
							{} 
						\end{ytableau} \;
					)_L^{ ( -\eta_{\bm{35}} , -\eta'_{\bm{35}} ) }
					\oplus
					\qty( \;
						\begin{ytableau}
							{} \\
							{}
						\end{ytableau} \; , \quad
						\begin{ytableau}
							{} & a \\
							{}
						\end{ytableau} \; , \quad
						\begin{ytableau}
							{} 
						\end{ytableau} \;
					)_R^{ (\eta_{\bm{35}} , \eta'_{\bm{35}} ) } 
					\oplus \cdots \; .
				\end{aligned}
		\end{equation*}
		One might think that the ${\bm{35}}$ can be allowed by the condition ($\star$), 
		if we choose the $Z_2$ parity as $(\eta_{\bm{35}} , \eta'_{\bm{35}} ) = (+,+)$ or $(-,-)$.
		However, 
		extra massless exotic fermions $(\bm{3}, \bm{2})$ or $(\bm{3}^*, \bm{2})$ appears in either choice.
		Therefore, $\bm{35}$ is not suitable, too.
		After all, we conclude that the representations satisfying the condition ($\star$) are only 
		\begin{equation*}
			\mathcal{R} =
				\bm{6}\, , \, \bm{21}
				= 
				\ytableausetup{ boxsize = normal, centertableaux}
				\begin{ytableau}
					{} 
				\end{ytableau}
				\; , \;
				\begin{ytableau}
					{} & {} 
				\end{ytableau}
				\, .
		\end{equation*}
		 ( (\ref{representations}) in Section 3 shows these validities.)
	\item $ n = 0, \, m \geq 1 $ . \\
		In this case, we can write $\mathcal{R}$ as 
		\begin{equation*}
		\ytableausetup{ boxsize = normal }
			\mathcal{R} = \;
				\begin{aligned}
					&\overbrace{\hspace{4.45em}}^j \overbrace{\hspace{4.45em}}^k 
					\overbrace{\hspace{4.45em}}^l \overbrace{\hspace{6em}}^m \\[-2mm]
					&\begin{ytableau}
						{} & \cdots & {} & {} & \cdots & {} & {} & \cdots & {} & {} & \cdots & {} & {} \\
						{} & \cdots & {} & {} & \cdots & {} & {} & \cdots & {} & {} & \cdots & {} & a \\
						{} & \cdots & {} & {} & \cdots & {} & {} & \cdots & {} \\
						{} & \cdots & {} & {} & \cdots & {} \\
						{} & \cdots & {} \\
					\end{ytableau}
				\end{aligned}
				\, .
		\end{equation*}
		Decomposing $\mathcal{R}$ into representations of $SU(3)_c \times SU(2)_L \times U(1)_Y \times U(1)_X$ and
		focusing on blocks as
		\begin{equation*}
			\begin{aligned}
				&\begin{ytableau}
					{} & \cdots & {} & {} & \cdots & {} & {} & \cdots & {} & {} & \cdots & {} & {} \\
					{} & \cdots & {} & {} & \cdots & {} & {} & \cdots & {} & {} & \cdots & {} & \none & \none & a \\
					\none \\
					{} & \cdots & {} & {} & \cdots & {} & {} & \cdots & {} \\
					{} & \cdots & {} & {} & \cdots & {} \\
					\none \\
					{} & \cdots & {} \\
				\end{ytableau}
			\end{aligned}
			\, ,
		\end{equation*}
		one finds that $\mathcal{R}$ contains fields $ \psi_3, \, \psi_4 $ with the following representations
		\begin{equation*}
			\begin{aligned}
			\ytableausetup{ boxsize = 1.3em }
				&\psi_3 : \qty( \; 
					\raisebox{3mm}{ {\footnotesize$
					\begin{aligned}
						&\overbrace{\hspace{3.85em}}^j \overbrace{\hspace{3.85em}}^k 
						\overbrace{\hspace{3.85em}}^l \overbrace{\hspace{5.2em}}^m \\[-2mm]
						&\begin{ytableau}
							{} & \cdots & {} & {} & \cdots & {} & {} & \cdots & {} & {} & \cdots & {} & {} \\
							{} & \cdots & {} & {} & \cdots & {} & {} & \cdots & {} & {} & \cdots & {} & a
						\end{ytableau}
					\end{aligned} $}} \raisebox{-4mm}{,} \: 
					\raisebox{3mm}{ {\footnotesize$
					\begin{aligned}
						&\overbrace{\hspace{3.8em}}^j \overbrace{\hspace{3.8em}}^k 
						\overbrace{\hspace{3.8em}}^l \\[-2mm]
						&\begin{ytableau}
							{} & \cdots & {} & {} & \cdots & {} & {} & \cdots & {} \\
							{} & \cdots & {} & {} & \cdots & {} 
						\end{ytableau}
					\end{aligned} $}} \raisebox{-4mm}{,} \: 
					\raisebox{3mm}{ {\footnotesize$
					\begin{aligned}
						&\overbrace{\hspace{3.8em}}^j \\[-2mm]
						&\begin{ytableau}
							{} & \cdots & {} 
						\end{ytableau}
					\end{aligned} \; $}} ) \, , \\
				&\psi_4 : \qty( \; 
					\raisebox{3mm}{ {\footnotesize$
					\begin{aligned}
						&\overbrace{\hspace{3.85em}}^j \overbrace{\hspace{3.85em}}^k 
						\overbrace{\hspace{3.85em}}^l \overbrace{\hspace{5.2em}}^m \\[-2mm]
						&\begin{ytableau}
							{} & \cdots & {} & {} & \cdots & {} & {} & \cdots & {} & {} & \cdots & {} & {} \\
							{} & \cdots & {} & {} & \cdots & {} & {} & \cdots & {} & {} & \cdots & {} 
						\end{ytableau}
					\end{aligned} $}} \raisebox{-4mm}{,} \: 
					\raisebox{3mm}{ {\footnotesize$
					\begin{aligned}
						&\overbrace{\hspace{3.85em}}^j \overbrace{\hspace{3.85em}}^k 
						\overbrace{\hspace{5.2em}}^{l + 1} \\[-2mm]
						&\begin{ytableau}
							{} & \cdots & {} & {} & \cdots & {} & {} & \cdots & {} & a \\
							{} & \cdots & {} & {} & \cdots & {} 
						\end{ytableau}
					\end{aligned} $}} \raisebox{-4mm}{,} \: 
					\raisebox{3mm}{ {\footnotesize$
					\begin{aligned}
						&\overbrace{\hspace{3.85em}}^j \\[-2mm]
						&\begin{ytableau}
							{} & \cdots & {} 
						\end{ytableau}
					\end{aligned} \; $}} ) \, . \\
			\end{aligned}
		\end{equation*}
		$Z_2$ parities $(P, P')$ of ${\psi_3}_L \, , \, {\psi_3}_R \, ,\, {\psi_4}_L $ and ${\psi_4}_R$ are given as
		\begin{equation}
			\begin{alignedat}{3}
				{\psi_3}_L :\, &\big( &   &(-1)^j \eta_{\mathcal{R}}, &	  &(-1)^l \eta_{\mathcal{R}}' \, \big) \, ,\\
				{\psi_3}_R :\, &\big( &  -&(-1)^j \eta_{\mathcal{R}}, & \; -&(-1)^l \eta_{\mathcal{R}}' \, \big) \, ,\\
				{\psi_4}_L :\, &\big( &   &(-1)^j \eta_{\mathcal{R}}, & 	 -&(-1)^l \eta_{\mathcal{R}}' \, \big) \, ,\\
				{\psi_4}_R :\, &\big( & -&(-1)^j \eta_{\mathcal{R}}, &  	  &(-1)^l \eta_{\mathcal{R}}' \, \big) \, .
			\end{alignedat}
		\end{equation}
		Therefore, we can find the following properties
		\begin{enumerate}
			\item If $ m \geq 2$,  $\mathcal{R}$ is not consistent with the condition $(\star)$.
			\item If $ m = 1 $, only when  $j = k = l = 0$, $\mathcal{R}$ is consistent with the condition ($\star$),
		\end{enumerate}
		which tells us that the representations satisfying the condition ($\star$) are only
		\begin{equation}
			\mathcal{R} = 
				\bm{15} =
				\ytableausetup{ boxsize = normal, centertableaux}
				\begin{ytableau}
					{} \\
					{}
				\end{ytableau} 
				\, .
		\end{equation}
	\item $ n = m = 0, \, l \geq 1 $ .\\
		In this case, we can write $\mathcal{R}$ as 
		\begin{equation*}
			\mathcal{R} = \;
				\begin{aligned}
					&\overbrace{\hspace{4.45em}}^j \overbrace{\hspace{4.45em}}^k 
					\overbrace{\hspace{6em}}^l \\[-2mm]
					&\begin{ytableau}
						{} & \cdots & {} & {} & \cdots & {} & {} & \cdots & {} & {} \\
						{} & \cdots & {} & {} & \cdots & {} & {} & \cdots & {} & a \\
						{} & \cdots & {} & {} & \cdots & {} & {} & \cdots & {} & b \\
						{} & \cdots & {} & {} & \cdots & {} \\
						{} & \cdots & {} \\
					\end{ytableau}
					\, .
				\end{aligned}
		\end{equation*}
		Decomposing $\mathcal{R}$ into representations of $SU(3)_c \times SU(2)_L \times U(1)_Y \times U(1)_X$ and
		focusing on blocks as
		\begin{equation*}
			\begin{aligned}
				&\begin{ytableau}
					{} & \cdots & {} & {} & \cdots & {} & {} & \cdots & {} & {} \\
					{} & \cdots & {} & {} & \cdots & {} & {} & \cdots & {} & \none & \none & a \\
					\none & \none & \none & \none & \none & \none & \none & \none & \none & \none & \none & b \\
					{} & \cdots & {} & {} & \cdots & {} & {} & \cdots & {}  \\
					{} & \cdots & {} & {} & \cdots & {} \\
					\none \\
					{} & \cdots & {} \\
				\end{ytableau}
			\end{aligned}
			\, ,
		\end{equation*}
		one finds that $\mathcal{R}$ contains fields $ \psi_5, \, \psi_6 $ with the following representations
		\begin{equation*}
			\begin{aligned}
			\ytableausetup{ boxsize = 1.3em }
				&\psi_5 : \qty( \; 
					\raisebox{3mm}{ $
					\begin{aligned}
						&\overbrace{\hspace{3.85em}}^j \overbrace{\hspace{3.85em}}^k 
						\overbrace{\hspace{5.2em}}^l \\[-2mm]
						&\begin{ytableau}
							{} & \cdots & {} & {} & \cdots & {} & {} & \cdots & {} & {} \\
							{} & \cdots & {} & {} & \cdots & {} & {} & \cdots & {} & a
						\end{ytableau}
					\end{aligned} $} \raisebox{-4mm}{,} \: 
					\raisebox{3mm}{ $
					\begin{aligned}
						&\overbrace{\hspace{3.8em}}^j \overbrace{\hspace{3.8em}}^k 
						\overbrace{\hspace{5.2em}}^l \\[-2mm]
						&\begin{ytableau}
							{} & \cdots & {} & {} & \cdots & {} & {} & \cdots & {} & b \\
							{} & \cdots & {} & {} & \cdots & {} 
						\end{ytableau}
					\end{aligned} $} \raisebox{-4mm}{,} \: 
					\raisebox{3mm}{ $
					\begin{aligned}
						&\overbrace{\hspace{3.8em}}^j \\[-2mm]
						&\begin{ytableau}
							{} & \cdots & {} 
						\end{ytableau}
					\end{aligned} \; $} ) \, ,\\
				&\psi_6 : \qty( \; 
					\raisebox{3mm}{ $
					\begin{aligned}
						&\overbrace{\hspace{3.85em}}^j \overbrace{\hspace{3.85em}}^k 
						\overbrace{\hspace{5.2em}}^l \\[-2mm]
						&\begin{ytableau}
							{} & \cdots & {} & {} & \cdots & {} & {} & \cdots & {} & {} \\
							{} & \cdots & {} & {} & \cdots & {} & {} & \cdots & {} 
						\end{ytableau}
					\end{aligned} $} \raisebox{-4mm}{,} \: 
					\raisebox{3mm}{ $
					\begin{aligned}
						&\overbrace{\hspace{3.85em}}^j \overbrace{\hspace{3.85em}}^k 
						\overbrace{\hspace{5.2em}}^l \\[-2mm]
						&\begin{ytableau}
							{} & \cdots & {} & {} & \cdots & {} & {} & \cdots & {} &a \\
							{} & \cdots & {} & {} & \cdots & {} 
						\end{ytableau}
					\end{aligned} $} \raisebox{-4mm}{,} \: 
					\raisebox{3mm}{ $
					\begin{aligned}
						&\overbrace{\hspace{5.2em}}^{j + 1} \\[-2mm]
						&\begin{ytableau}
							{} & \cdots & {} & b
						\end{ytableau}
					\end{aligned} \; $} ) \, .\\
			\end{aligned}
		\end{equation*}
		$Z_2$ parities $(P, P')$ of ${\psi_5}_L \, , \, {\psi_5}_R \, ,\, {\psi_6}_L $ and ${\psi_6}_R$ are given as
		\begin{equation}
			\begin{alignedat}{3}
				{\psi_5}_L :\, &\big( &   &(-1)^j \eta_{\mathcal{R}}, &	  &(-1)^l \eta_{\mathcal{R}}' \, \big) \, ,\\
				{\psi_5}_R :\, &\big( &  -&(-1)^j \eta_{\mathcal{R}}, & \; -&(-1)^l \eta_{\mathcal{R}}' \, \big) \, ,\\
				{\psi_6}_L :\, &\big( &  -&(-1)^j \eta_{\mathcal{R}}, & 	  &(-1)^l \eta_{\mathcal{R}}' \, \big) \, ,\\
				{\psi_6}_R :\, &\big( &   &(-1)^j \eta_{\mathcal{R}}, &  	 -&(-1)^l \eta_{\mathcal{R}}' \, \big) \; .
			\end{alignedat}
		\end{equation}
		We can find the following properties.
		\begin{enumerate}
			\item If $ l \geq 2$,  $\mathcal{R}$ is not consistent with the condition $(\star)$.
			\item If $ l = 1 $, only when  $j = k = 0$, $\mathcal{R}$ is consistent with the condition ($\star$),
		\end{enumerate}
		which tells us that the representations satisfying the condition ($\star$) are only
		\begin{equation}
			\mathcal{R} = 
				\bm{20} =
				\ytableausetup{ boxsize = normal, centertableaux}
				\begin{ytableau}
					{} \\
					{} \\
					{}
				\end{ytableau} 
				\, .
		\end{equation}
	\item $ n = m = l = 0, \, k \geq 1 $ . \\
		In this case, we can write $\mathcal{R}$ as 
		\begin{equation*}
			\mathcal{R} = \;
				\begin{aligned}
					&\overbrace{\hspace{4.45em}}^j \overbrace{\hspace{6em}}^k \\[-2mm]
					&\begin{ytableau}
						{} & \cdots & {} & {} & \cdots & {} & {} \\
						{} & \cdots & {} & {} & \cdots & {} & a \\
						{} & \cdots & {} & {} & \cdots & {} & b \\
						{} & \cdots & {} & {} & \cdots & {} & c \\
						{} & \cdots & {} \\
					\end{ytableau}
				\end{aligned}
				\, .
		\end{equation*}
		Decomposing $\mathcal{R}$ into the representations of $SU(3)_c \times SU(2)_L \times U(1)_Y \times U(1)_X$ and
		focusing on blocks as
		\begin{equation*}
			\begin{aligned}
				&\begin{ytableau}
					{} & \cdots & {} & {} & \cdots & {} & {} \\
					{} & \cdots & {} & {} & \cdots & {} & \none & \none & a  \\
					\none & \none & \none & \none & \none & \none & \none & \none & b \\
					{} & \cdots & {} & {} & \cdots & {} & \none & \none & c \\
					{} & \cdots & {} & {} & \cdots & {} \\
					\none \\
					{} & \cdots & {} \\
				\end{ytableau}
			\end{aligned}
			\, ,
		\end{equation*}
		one finds that $\mathcal{R}$ contains fields $ \psi_7, \, \psi_8 $ with the following representations
		\begin{equation*}
			\begin{aligned}
			\ytableausetup{ boxsize = 1.3em }
				&\psi_7 : \qty( \; 
					\raisebox{3mm}{ $
					\begin{aligned}
						&\overbrace{\hspace{3.85em}}^j \overbrace{\hspace{5.2em}}^k \\[-2mm]
						&\begin{ytableau}
							{} & \cdots & {} & {} & \cdots & {} & {} \\
							{} & \cdots & {} & {} & \cdots & {} & a
						\end{ytableau}
					\end{aligned} $} \raisebox{-4mm}{,} \: 
					\raisebox{3mm}{ $
					\begin{aligned}
						&\overbrace{\hspace{3.8em}}^j \overbrace{\hspace{5.2em}}^k \\[-2mm]
						&\begin{ytableau}
							{} & \cdots & {} & {} & \cdots & {} & b \\
							{} & \cdots & {} & {} & \cdots & {} & c
						\end{ytableau}
					\end{aligned} $} \raisebox{-4mm}{,} \: 
					\raisebox{3mm}{ $
					\begin{aligned}
						&\overbrace{\hspace{3.8em}}^j \\[-2mm]
						&\begin{ytableau}
							{} & \cdots & {} 
						\end{ytableau}
					\end{aligned} \; $} ) \, ,\\
				&\psi_8 : \qty( \; 
					\raisebox{3mm}{ $
					\begin{aligned}
						&\overbrace{\hspace{3.85em}}^j \overbrace{\hspace{5.2em}}^k \\[-2mm]
						&\begin{ytableau}
							{} & \cdots & {} & {} & \cdots & {} & {} \\
							{} & \cdots & {} & {} & \cdots & {} 
						\end{ytableau}
					\end{aligned} $} \raisebox{-4mm}{,} \: 
					\raisebox{3mm}{ $
					\begin{aligned}
						&\overbrace{\hspace{3.85em}}^j \overbrace{\hspace{5.2em}}^k \\[-2mm]
						&\begin{ytableau}
							{} & \cdots & {} & {} & \cdots & {} & a \\
							{} & \cdots & {} & {} & \cdots & {} & b
						\end{ytableau}
					\end{aligned} $} \raisebox{-4mm}{,} \: 
					\raisebox{3mm}{ $
					\begin{aligned}
						&\overbrace{\hspace{5.2em}}^{j + 1} \\[-2mm]
						&\begin{ytableau}
							{} & \cdots & {} & c
						\end{ytableau}
					\end{aligned} \; $} ) \, . \\
			\end{aligned}
		\end{equation*}
		$Z_2$ parities $(P, P')$ of ${\psi_7}_L \, , \, {\psi_7}_R \, ,\, {\psi_8}_L $ and ${\psi_8}_R$ are given as
		\begin{equation}
			\begin{alignedat}{3}
				{\psi_7}_L :\, &\big( &   &(-1)^j \eta_{\mathcal{R}}, &	  &\eta_{\mathcal{R}}' \, \big) \, ,\\
				{\psi_7}_R :\, &\big( &  -&(-1)^j \eta_{\mathcal{R}}, & \; -&\eta_{\mathcal{R}}' \, \big) \, ,\\
				{\psi_8}_L :\, &\big( &  -&(-1)^j \eta_{\mathcal{R}}, & 	  &\eta_{\mathcal{R}}' \, \big) \, ,\\
				{\psi_8}_R :\, &\big( &   &(-1)^j \eta_{\mathcal{R}}, &  	  -&\eta_{\mathcal{R}}' \, \big)\, .
			\end{alignedat}
		\end{equation}
		We can find the following properties.
		\begin{enumerate}
			\item If $ k \geq 2$,  $\mathcal{R}$ is not consistent with the condition $(\star)$.
			\item If $ k = 1 $, only when  $j = 0$, $\mathcal{R}$ is consistent with the condition ($\star$),
		\end{enumerate}
		which tells us that the representations satisfying the condition ($\star$) are only
		\begin{equation}
			\mathcal{R} = 
				\bm{15}^* =
				\ytableausetup{ boxsize = normal, centertableaux}
				\begin{ytableau}
					{} \\
					{} \\
					{} \\
					{}
				\end{ytableau} 
				\, .
		\end{equation}
		in this case.
	\item $ n = m = l = k = 0, \, j \geq 1$ .\\
		In this case, we can write $\mathcal{R}$ as 
		\begin{equation*}
			\mathcal{R} = \;
				\begin{aligned}
					&\overbrace{\hspace{6em}}^j \\[-2mm]
					&\begin{ytableau}
						{} & \cdots & {} & {} \\
						{} & \cdots & {} & {} \\
						{} & \cdots & {} & a \\
						{} & \cdots & {} & b \\
						{} & \cdots & {} & {}
					\end{ytableau}
				\end{aligned}
				\, .
		\end{equation*}
		Decomposing $\mathcal{R}$ into representations of $SU(3)_c \times SU(2)_L \times U(1)_Y \times U(1)_X$ and
		focusing on blocks as
		\begin{equation*}
			\begin{aligned}
				&\begin{ytableau}
					{} & \cdots & {} & {} \\
					{} & \cdots & {} & {} \\
					{} & \cdots & {} & \none & \none & a \\
					\none & \none & \none & \none & \none & b \\
					{} & \cdots & {} \\
					\none \\
					{} & \cdots & {} & {} \\
				\end{ytableau}
			\end{aligned}
			\, ,
		\end{equation*}
		one finds that $\mathcal{R}$ contains fields $ \psi_9, \, \psi_{10} $ with the following representations
		\begin{equation*}
			\begin{aligned}
			\ytableausetup{ boxsize = 1.3em }
				&\psi_9 : \qty( \; 
					\raisebox{3mm}{ $
					\begin{aligned}
						&\overbrace{\hspace{5.2em}}^j \\[-2mm]
						&\begin{ytableau}
							{} & \cdots & {} & {} \\
							{} & \cdots & {} & {} \\
							{} & \cdots & {} & a
						\end{ytableau}
					\end{aligned} $} \raisebox{-4mm}{,} \: 
					\raisebox{3mm}{ $
					\begin{aligned}
						&\overbrace{\hspace{5.2em}}^j \\[-2mm]
						&\begin{ytableau}
							{} & \cdots & {} & b
						\end{ytableau}
					\end{aligned} $} \raisebox{-4mm}{,} \: 
					\raisebox{3mm}{ $
					\begin{aligned}
						&\overbrace{\hspace{5.2em}}^j \\[-2mm]
						&\begin{ytableau}
							{} & \cdots & {} & {}
						\end{ytableau}
					\end{aligned} \; $} ) \, ,\\
				&\psi_{10} : \qty( \; 
					\raisebox{3mm}{ $
					\begin{aligned}
						&\overbrace{\hspace{5.2em}}^j \\[-2mm]
						&\begin{ytableau}
							{} & \cdots & {} & {} \\
							{} & \cdots & {} & {} \\
							{} & \cdots & {} 
						\end{ytableau}
					\end{aligned} $} \raisebox{-4mm}{,} \: 
					\raisebox{3mm}{ $
					\begin{aligned}
						&\overbrace{\hspace{5.2em}}^j \\[-2mm]
						&\begin{ytableau}
							{} & \cdots & {} & a \\
							b
						\end{ytableau}
					\end{aligned} $} \raisebox{-4mm}{,} \: 
					\raisebox{3mm}{ $
					\begin{aligned}
						&\overbrace{\hspace{5.2 em}}^j \\[-2mm]
						&\begin{ytableau}
							{} & \cdots & {} & {}
						\end{ytableau}
					\end{aligned} \; $} ) \, .\\
			\end{aligned}
		\end{equation*}
		$Z_2$ parities $(P, P')$ of ${\psi_9}_L \, , \, {\psi_9}_R \, ,\, {\psi_{10}}_L $ and ${\psi_{10}}_R$ are given as
		\begin{equation}
			\begin{alignedat}{3}
				{\psi_9}_L :\, &\big( &   &(-1)^j \eta_{\mathcal{R}}, &	  &(-1)^j \eta_{\mathcal{R}}' \, \big) \, ,\\
				{\psi_9}_R :\, &\big( &  -&(-1)^j \eta_{\mathcal{R}}, & \; -&(-1)^j \eta_{\mathcal{R}}' \, \big) \, ,\\
				{\psi_{10}}_L :\, &\big( &   &(-1)^j \eta_{\mathcal{R}}, &  -&(-1)^j \eta_{\mathcal{R}}' \, \big) \, ,\\
				{\psi_{10}}_R :\, &\big( & -&(-1)^j \eta_{\mathcal{R}}, &    &(-1)^j \eta_{\mathcal{R}}' \, \big) \, .
			\end{alignedat}
		\end{equation}
		Therefore, looking at representations of $SU(2)$, we can say
		\begin{enumerate}
			\item If $ j \geq 3$,  $\mathcal{R}$ is not consistent with the condition $(\star)$,
		\end{enumerate}
		which means that the representations satisfying the condition ($\star$) are only
		\begin{equation}
			\mathcal{R} = 
				\bm{6}^* ,\, \bm{21}^* =
				\ytableausetup{ boxsize = normal, centertableaux}
				\begin{ytableau}
					{} \\
					{} \\
					{} \\
					{} \\
					{}
				\end{ytableau} \; , \;
				\begin{ytableau}
					{} & {} \\
					{} & {} \\
					{} & {} \\
					{} & {} \\
					{} & {}
				\end{ytableau} \, .
		\end{equation}
\end{enumerate}
Summarizing all of the above results together, 
we conclude that the representations satisfying the condition ($\star$) are obtained as
\begin{equation}
	\mathcal{R} = \bm{6}, \bm{6}^*, \bm{15}, \bm{15}^*, \bm{20}, \bm{21}, \bm{21}^*.
\end{equation}


\vspace*{0.5cm}

{}

\end{document}